\documentclass[aps,prc,twocolumn,amsmath,amssymb,superscriptaddress,showpacs,showkeys,preprintnumbers,nofootinbib]{revtex4-2}
\usepackage{dcolumn}
\usepackage{graphicx,color}
\usepackage{multirow}
\usepackage{textcomp}
\usepackage{subfigure}
\usepackage{physics}
\usepackage{dsfont}
\usepackage{comment}
\usepackage{relsize}

\usepackage{tikz}	
\begin{document}
	\newcommand {\nc} {\newcommand}
	\nc {\beq} {\begin{eqnarray}}
	\nc {\eeq} {\nonumber \end{eqnarray}}
	\nc {\eeqn}[1] {\label {#1} \end{eqnarray}}
\nc {\eol} {\nonumber \\}
\nc {\eoln}[1] {\label {#1} \\}
\nc {\ve} [1] {\mbox{\boldmath $#1$}}
\nc {\ves} [1] {\mbox{\boldmath ${\scriptstyle #1}$}}
\nc {\mrm} [1] {\mathrm{#1}}
\nc {\half} {\mbox{$\frac{1}{2}$}}
\nc {\thal} {\mbox{$\frac{3}{2}$}}
\nc {\fial} {\mbox{$\frac{5}{2}$}}
\nc {\la} {\mbox{$\langle$}}
\nc {\ra} {\mbox{$\rangle$}}
\nc {\etal} {\emph{et al.}}
\nc {\eq} [1] {(\ref{#1})}
\nc {\Eq} [1] {Eq.~(\ref{#1})}
\nc {\Refc} [2] {Refs.~\cite[#1]{#2}}
\nc {\Sec} [1] {Sec.~\ref{#1}}
\nc {\chap} [1] {Chapter~\ref{#1}}
\nc {\anx} [1] {Appendix~\ref{#1}}
\nc {\tbl} [1] {Table~\ref{#1}}
\nc {\Fig} [1] {Fig.~\ref{#1}}
\nc {\ex} [1] {$^{#1}$}
\nc {\Sch} {Schr\"odinger }
\nc {\flim} [2] {\mathop{\longrightarrow}\limits_{{#1}\rightarrow{#2}}}
\nc {\textdegr}{$^{\circ}$}
\nc {\inred} [1]{\textcolor{red}{#1}}
\nc {\inblue} [1]{\textcolor{blue}{#1}}
\nc {\IR} [1]{\textcolor{red}{#1}}
\nc {\IB} [1]{\textcolor{blue}{#1}}
\nc{\pderiv}[2]{\cfrac{\partial #1}{\partial #2}}
\nc{\deriv}[2]{\cfrac{d#1}{d#2}}

\nc {\bit} {\begin{itemize}}
	\nc {\eit} {\end{itemize}}

\title{Uncertainty quantification in  $(p,n)$ reactions}
\author{A. J. Smith}
\email{smithan@frib.msu.edu}
\affiliation{Facility for Rare Isotope Beams, Michigan State University, East Lansing, Michigan 48824, USA}
\affiliation{Department of Physics and Astronomy, Michigan State University, East Lansing, Michigan 48824, USA}
\author{C.~Hebborn}
\email{hebborn@frib.msu.edu}
\affiliation{Facility for Rare Isotope Beams, Michigan State University, East Lansing, Michigan 48824, USA}
\affiliation{Department of Physics and Astronomy, Michigan State University, East Lansing, Michigan 48824, USA}
\author{F.~M.~Nunes}
\affiliation{Facility for Rare Isotope Beams, Michigan State University, East Lansing, Michigan 48824, USA}
\affiliation{Department of Physics and Astronomy, Michigan State University, East Lansing, Michigan 48824, USA}
\author{R. G. T. Zegers}
\affiliation{Facility for Rare Isotope Beams, Michigan State University, East Lansing, Michigan 48824, USA}
\affiliation{Department of Physics and Astronomy, Michigan State University, East Lansing, Michigan 48824, USA}

\date{\today}
\begin{abstract}
Charge-exchange reactions are versatile probes for  nuclear structure. In particular, when populating isobaric analog states, these reactions are used to study  isovector nuclear densities and  neutron skins. The quality of the information extracted from charge-exchange data depends on the accuracy of the reaction models and their inputs; this work addresses  these two points. First, we quantify the uncertainties due to effective nucleon-nucleus interactions by propagating the parameter posterior distributions of the recent global optical model KDUQ~\cite{KDUQ} to $(p,n)$ reaction observables populating the isobaric analogue state, at beam energies in the range of $25-160$~MeV.  Our analysis, focusing on $^{48}$Ca, shows that  the total parametric uncertainties on the cross sections are around 60-100\%. The source of this uncertainty is mainly  the transition operator as the uncertainties from the distorted waves alone are less than about 15\%. Second, we  perform a comparison between  two- and three-body models that both describe the dynamics of the reaction within the DWBA. The predictions from these two  models are similar  and generally agree with the available data, suggesting that 1-step DWBA is sufficient to describe the reaction process. Only at a beam energy of 25 MeV there are possibly signs that a 1-step assumption is not fully correct.   
This work  provides motivation for the quantification of uncertainties associated with the transition operator in three-body model. It also suggests that further constraint of the optical potential parameters is needed for increased model precision.
\end{abstract}

\maketitle

\section{Introduction}

One of the most exciting open questions in our field today concerns the equation of state of nuclear matter as it evolves from symmetric nuclear matter to neutron rich matter \cite{review-eos}. The key element in this equation of state is the symmetry energy, a term that relates directly to neutron/proton asymmetry in the system. The quest to constrain the symmetry energy encompasses a variety of projects  from  theory, experiment and observations (e.g. Refs.~\cite{brown2013,drischler2020,central2019,duncan2023,sorenson2024}).

One way to explore the evolution of nuclear structure with neutron/proton asymmetry is through charge-exchange reactions. In particular, charge-exchange reactions populating the isobaric analogue state (IAS) of the original nucleus through a $\Delta S=0$, $\Delta L=0$ Fermi transition, probe the isovector component of the transition operator. It has long been understood that this component of the mean field is responsible for phenomena such as neutron skins \cite{brown2013,Loc2014}, and that it
provides a direct handle on the symmetry energy around nuclear densities. For example, in Ref.~\cite{danielewicz2017} Danielewicz and collaborators studied charge-exchange reactions to the IAS and obtained information on the isovector component of the optical potential, which in turn informed the symmetry energy. While there are many types of charge-exchange reactions that are used for a variety of studies of the isovector response of nuclei, in this work we focus solely on $(p,n)$ reactions  that populate the IAS.

The interpretation of $(p,n)$ charge-exchange measurements relies on a reaction model. There are two main approaches available in the field, one that is based on a two-body theory (sometimes referred to as the macroscopic approach) \cite{danielewicz2017} and another one that is based on a three-body framework (often referred to as a microscopic approach)~\cite{Udagawa1987,Taddeucci1987}. The latter approach is most commonly used for interpretations of the isovector response of charge-exchange reactions. For the purpose of studying the IAS, both approaches are adopted and the two-body approach provides a direct and simple connection to the isovector component of the optical potential. In either case, the inputs are  effective potentials to describe the initial and final distorted waves and the transition operator, all of which carry uncertainties. 

Recent work \cite{whiteheadchex}  assessed the magnitude of  uncertainties in the two-body model for specific charge-exchange reactions, namely $(p,n)$ reactions on $^{14}$C, $^{48}$Ca, and $^{90}$Zr  at beam energies $E=25, 35, 45$ MeV. In that study, neutron and proton elastic-scattering data were independently used to constrain the parameters of the relevant proton- and neutron-target optical potentials, and from the parameter posterior distributions obtained from the neutron and proton separate Bayesian calibrations, credible intervals for the charge-exchange cross sections were determined. The errors obtained were surprisingly large, due to the assumption that the neutron and proton optical potentials were assumed to be independent. In reality, these potentials are related and their respective parameters are strongly correlated.

In this work, we revisit the uncertainties in charge-exchange reactions to IAS using a global optical potential for which full Bayesian uncertainty estimation has been performed (KDUQ) \cite{KDUQ}. Because KDUQ is a global parameterization, including a wide range of neutron and proton data on mostly stable nuclei, the correlation between neutron and proton optical potentials are intrinsic by construction. Also included in KDUQ is the energy dependence  across a wide range of beam energies. In this way, KDUQ provides an excellent starting point to study the effect of optical model uncertainties in charge-exchange observables.

This work expands the study of Ref.~\cite{whiteheadchex} to include a comparison between the two-body and the three-body models. Preliminary steps toward such a comparison was done in Ref.~\cite{TPPthesis}, however this work is the first study in which a full comparison is performed including uncertainty quantification. Finally, we also expand the energy range of the reactions to include energies up to $E=160$~MeV, given the potential interest of exploring these reactions on rare isotopes produced in fragmentation facilities such as RIKEN and FRIB. We focus on the $^{48}$Ca($p$,$n$) reaction, populating the IAS in $^{48}$Sc.

The paper is organized in the following way: Section II provides a brief summary of the theoretical framework for both the two-body and three-body approaches; in Section~III we present the results; and conclusions are drawn in Section IV.

\section{Theoretical framework}

In this work, we focus on $A(p,n)B$ charge-exchange reactions populating the isobaric analog state $B$ of the nucleus $A$.
The  charge-exchange cross section can be obtained from the corresponding T-matrix  which in the distorted wave born approximation (DWBA) reads 
\begin{equation}
T^{{\rm DWBA}} = \bra{\psi_f}\hat O \ket{\psi_i} \label{eq1}
\end{equation}
where $\hat O$ is the transition operator, and  $\ket{\psi_i}$, and $\ket{\psi_f}$ are the incoming $A$-$p$ and outgoing $B$-$n$ distorted waves, respectively.  In principle, the operator $\hat O$ and the scattering waves are many-body objects. However, such many-body descriptions are unfeasible for reactions  involving medium- and heavy-mass nuclei and energies above a few MeV.  In this work,   we simplify this many-body problem into a few-body one and we adopt  a two- or three-body description. 

In the two-body model,  internal degrees of freedom of the  nuclei in the entrance ($A$) and exit ($B$) channels  are not explicitly included as opposed to the three-body approach where these nuclei have single-particle structure. 
This few-body simplification comes at a cost, since the interactions between the nucleon projectile and the target cluster(s) are simulated through optical potentials, which includes an imaginary part modeling effectively the inelastic channels~\cite{feshbach1958unified,TN09,OpticalPotentialReview}.
In both the two- and three-body models, the scattering waves are obtained from $p$-$A$ and $n$-$B$ optical potentials, respectively $\hat U_{pA}$ and $\hat  U_{nB}$. The main difference between the two- and three-body models lies in the construction of the operator. {It is noted that in both models the transition strengths are equal and exhaust the Fermi sum rule \cite{IKEDA1963271}, $B(F)=(N-Z)=8$ for the case of $^{48}$Ca studied here.}   

\subsection{Two-body charge-exchange model}
   In the case of  $(p,n)$ charge-exchange reactions to the isobaric analog state, the two-body transition operator is directly related to the isovector part of these optical potentials
\begin{equation}
\hat O_{2B} = \frac{\sqrt{|N-Z|}}{N-Z-1} \left[\hat U_{nB} - \hat U_{pA} \right] \label{eq2}
\end{equation}
where $N$ is the number of neutrons, and $Z$ is the number of protons in the nucleus $A$. Because this two-body operator depends on the difference of optical potentials, its uncertainties are strongly reduced if optical potentials are expressed in terms of  isoscalar and isovector components~\cite{whiteheadchex}. The two-body calculations are performed with the code \textsc{CHEX} \cite{whiteheadchex}

\subsection{Three-body charge-exchange model}
\label{section:3B}
In the three-body model, the operator $\hat O_{3B}$ is calculated as a form factor, in which a suitable nucleon-nucleon ($NN$) interaction $T_{NN}$ is folded over the transition densities of the target-residue and projectile-ejectile system. In the case of the ($p$,$n$) reaction, a single folding over the transition density of the target-residue system, $\rho_{AB}$, is required \cite{Taddeucci1987, RevModPhys.64.491,Ichimura:2006}
\begin{equation}
\hat O_{3B} \propto \int dr  \,\rho_{AB}\,   T_{NN} \label{eq3}
\end{equation}
The transition densities $\rho_{AB}$ can be calculated in various models, but for light and medium-heavy nuclei, shell-model calculations are often used. For the excitation of the IAS of the $^{48}$Ca($0^{+}$) ground state in $^{48}$Sc, one-body transition densities were calculated in the $pf$ shell-model space with the \textsc{gx1a} interaction \cite{hon02,hon04,hon05} using the code \textsc{NuShellX} \cite{BROWN2014115}. The excitation of the IAS is almost completely due to $0f_{7/2}$-$0f_{7/2}$ proton-particle, neutron-hole excitations. 

The nucleon-nucleon interaction most commonly used, as is the case in this work, is the Love-Franey $T_{NN}$ matrix \cite{PhysRevC.24.1073,franey1985}, since it has the attractive feature that the different components of the interaction can easily be attributed to specific properties of a reaction, such as spin and isospin transfer. In addition, the Love-Franey interaction has proven to accurately represent key features of charge-exchange reactions, see e.g. \cite{Taddeucci1987, RevModPhys.64.491,Ichimura:2006}. The $T_{NN}$ matrix was originally developed for beam energies of 50 MeV and above. Recently, new parameterizations for beam energies of 50 MeV and below became available \cite{CAPPUZZELLO2023103999,lenske2023}. Since in this work we study ($p$,$n$) charge-exchange reactions for $E_{p}=25$ MeV to $E_{p}=160$ MeV, both sets of interactions are used.  

The calculated form factor is inserted in a DWBA calculation (Eq.1) to extract the differential cross sections. The three-body calculations were performed with the code \textsc{DW81} \cite{DW81}. It is noted that the inclusion of knock-on exchange amplitudes, which interfere with the direct amplitudes, is necessary when using the $T_{NN}$ matrix. In the calculations presented in this work with the code \textsc{DW81}, the treatment of these exchange amplitudes was done exactly, rather than using the a short-range approximation usually used for reactions with composite particles \cite{PhysRevC.24.1073,franey1985}.

\subsection{Optical Potential Parameters}
For the $p$-$^{48}$Ca and $n$-$^{48}$Sc optical potentials needed in both few-body models, we use the global optical  potentials fit by Koning and Delaroche (KD) and its recent version with quantified uncertainties (KDUQ), which are valid for  $ 24 \leq A \leq 209$ and 1~keV~$\leq E \leq 200 $ MeV. These two global parametrizations are both expressed in terms of isoscalar and isovector components, which have been fitted on the same corpus of data, including elastic scattering angular distributions and analyzing power, as well as total and reaction cross section data for stable nuclei.  The KDUQ parameter posterior distributions and their correlations were built using a Bayesian framework. In this work, we use the 416 samples provided in Ref.~\cite{KDUQ} to evaluate the parametric uncertainties in charge-exchange cross sections. 

\section{Results}

In this section, we aim to quantify the uncertainties in the differential charge-exchange cross sections due to the underlying parametric uncertainty in the optical model potential. Let us clarify that these uncertainties represent a lower bound as model uncertainties resulting from the simplified treatment of the DWBA have not been quantified in this analysis.

\subsection{Full parametric uncertainty quantification in the two-body model}

\begin{figure}
    \centering

    \includegraphics[width=\linewidth]{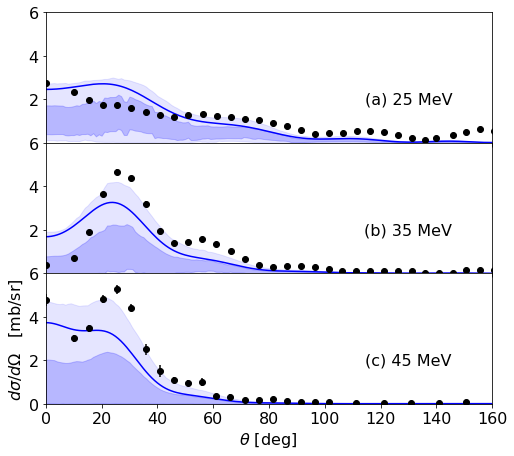}
  \includegraphics[width=\linewidth]{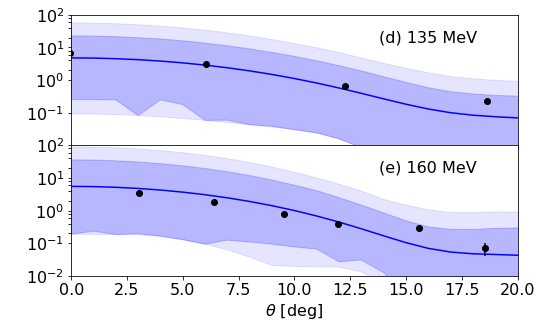}

        \caption{Charge-exchange cross sections  $^{48}$Ca$(p,n)^{48}$Sc($0^+$;IAS) at  (a) 25 MeV, (b) 35 MeV, (c)~45 MeV, (d) 135 MeV and (e) 160 MeV calculated within the two-body framework~\eqref{eq2}. The dark and light shaded blue bands correspond to the 68\% and 95\% credible intervals, obtained by propagating the KDUQ posteriors~\cite{KDUQ} consistently to the two-body operator and the scattering waves \eqref{eq2}. The solid blue  line represents the cross section computed from the original Koning-Delaroche parametrization. The data  (black points) were taken from Ref. \cite{data} }
    \label{fig:fig1}
    \end{figure}

We first perform a full quantification of parametric uncertainties,  associated with both the distorted waves and the operator, for  $^{48}$Ca$(p,n)^{48}$Sc($0^+$;IAS) at 25, 35, 45, 135, and 160 MeV (Fig.~\ref{fig:fig1}). Such complete uncertainty quantification can only be done consistently in the two-body framework, in which the scattering waves and the operator are  derived from the same optical potentials. The 68\% (dark shaded blue) and 95\% (light shaded blue) credible intervals, obtained by propagating the 416 samples of the KDUQ posteriors~\cite{KDUQ}, are compared to the cross section obtained with the original KD parametrization (solid blue line)~\cite{KD}. Although the predicted uncertainties are large, they are considerable smaller than those predicted in Ref.~\cite{whiteheadchex} due to the inclusion of the appropriate correlations between the neutron and proton optical potentials.

We now turn to the comparison between our predictions and the charge-exchange data (represented by the black dots in Fig. \ref{fig:fig1}).  At the lower energies [Fig. \ref{fig:fig1}(a,b,c)], the 1$\sigma$ credible intervals  do not capture  the data at the peak nor at backward angles. The difficulty in describing the angular distribution data for energies in the $E=25-45$ MeV range was  discussed in \cite{danielewicz2017,TPPthesis}. We show here, with uncertainty quantification, that agreement is obtained only at the $2\sigma$ level.
In contrast to the lower energy results, the model predictions reproduce the experimental angular distributions  within 1$\sigma$ for the higher beam energies  [Fig. \ref{fig:fig1}(d,e)] both in magnitude and in the dependence with angle.  This suggests that even at these higher energies,  the two-body model is effective. Nevertheless,  here again the widths of the credible intervals  are much larger than the experimental errors (around $100$\%).   In Table \ref{tab:table1}, we show the absolute and relative theoretical uncertainties at the peak of the charge-exchange angular distribution, corresponding to angle $\theta_{max}$\footnote{The absolute uncertainties at the peak of the cross section are defined between the limit of the uncertainty interval and its mean value. The relative ones are simply calculated from the absolute uncertainties normalized by the mean value.}. In the fourth column (‘‘Full UQ") we can see that the relative uncertainties increase from {62\%} at 25 MeV to {99\%} at 160 MeV. This enhancement at larger beam energies may be related to a larger sensitivity of charge-exchange observables to the volume of the interactions given that in KDUQ there is a larger  uncertainty in the volume integrals (see Fig. 12 of Ref.~\cite{KDUQ}).

\begin{table}
    \centering
    \begin{tabular}{cc||cc|cc||cc}
        \multirow{3.2}{*}{$E$}& \multirow{3.2}{*}{$\theta_{\text{max}}$} &\multicolumn{4}{c||}{Two-body}&\multicolumn{2}{c}{Three-body}\\\hline 
      &  & \multicolumn{2}{c|}{Full UQ} & \multicolumn{2}{c||}{UQ-DW}&\multicolumn{2}{c}{UQ-DW}\\       
   (MeV) & (deg.) & abs. & rel.  &abs. & rel.  &  abs. & rel. \\ \hline\hline 
        25 & 0 &  0.7 & {62\%}  & 0.3 & {13\%} & 0.30 & 15\%\\
        35 & 26 &  1.1 & {95\%}  & 0.4 &{11\%}& 0.47 & 11\%\\
        45 & 26 & 1.1 & {99\%}  &0.4 &{12\%}&  0.62 &13\%\\
        135 & 0 & {12.2} & {97\%}  &{0.5} &{10\%}&0.78&14\%\\
        160 & 0 & {17.8} & {99\%}  & 0.5 &10\%&0.49&10\%\\
    \end{tabular}
    \caption{Half width of the  68 $\%$  credible intervals at the peak $\theta_{\text{max}}$ (in degrees)  of the  $^{48}$Ca$(p,n)^{48}$Sc($0^{+}$;IAS) charge-exchange angular distributions (in mb/sr) for the various beam energies studied.  The absolute  (abs.) widths of these intervals and their relative (rel.) importance are given for the different cases: when the uncertainties in the operator and the distorted wave is considered (Full UQ, see Fig. \ref{fig:fig1}) and when only the uncertainties in the distorted waves are accounted for in both the two- and three-body reaction models (UQ-DW, see Fig.~\ref{fig:fig2}). }
    \label{tab:table1}
\end{table}

\begin{figure}
    \centering

        \includegraphics[trim=3cm 4.5cm 4cm 4.5cm, clip,width=\linewidth]{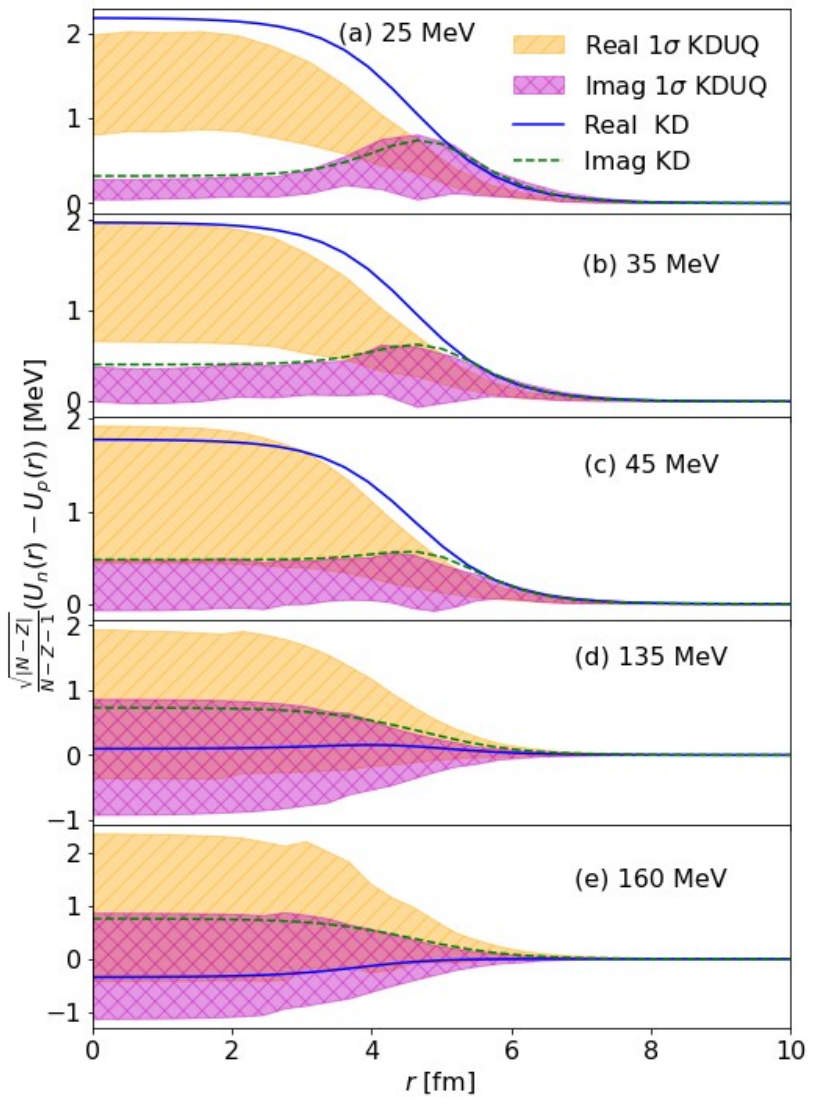}
        \caption{Real and imaginary part of the two-body transition operator in the $s$-wave for charge-exchange  reactions $^{48}$Ca$(p,n)^{48}$Sc($0^+$;IAS) at  (a) 25 MeV, (b) 35 MeV, (c)~45~MeV, (d) 135 MeV and (e) 160 MeV as a function of the nucleon-nucleus relative coordinate $r$.   The 68\% credible  intervals (shaded  forward    yellow and magenta  crossed hashmarks) are obtained from the KDUQ posteriors~\cite{KDUQ} while the solid  blue and dashed green  lines are calculated using the KD potential~\cite{KD}.   }        \label{fig:fig3}
\end{figure}

It may seem surprising that the charge-exchange cross section predictions with the original KD potential  do not lie within the $1\sigma$ intervals of KDUQ (solid blue  lines in Fig.~\ref{fig:fig1}). This is especially true for the three lowest energies. To inspect the reason for this mismatch, we provide in Fig.~\ref{fig:fig3} the operator $\hat O_{2B}$ responsible for the charge-exchange transition in the two-body framework: the real and imaginary parts of the original KD potential (solid blue and dashed green lines respectively) should be compared with real and imaginary KDUQ interval (the yellow forward  and pink crossed hashmarks respectively). Fig.~\ref{fig:fig3} shows that the operator obtained with the original KD potential falls just outside the operator 68\%  credible interval obtained from KDUQ for the lowest beam energies which explains the features seen in Fig.~\ref{fig:fig1}. This mismatch is no longer present for the highest energies where the uncertainty intervals become larger.

\begin{figure}
    \centering

        \includegraphics[trim=0.5cm 7cm 0.5cm 6.9cm, clip,width=\linewidth]{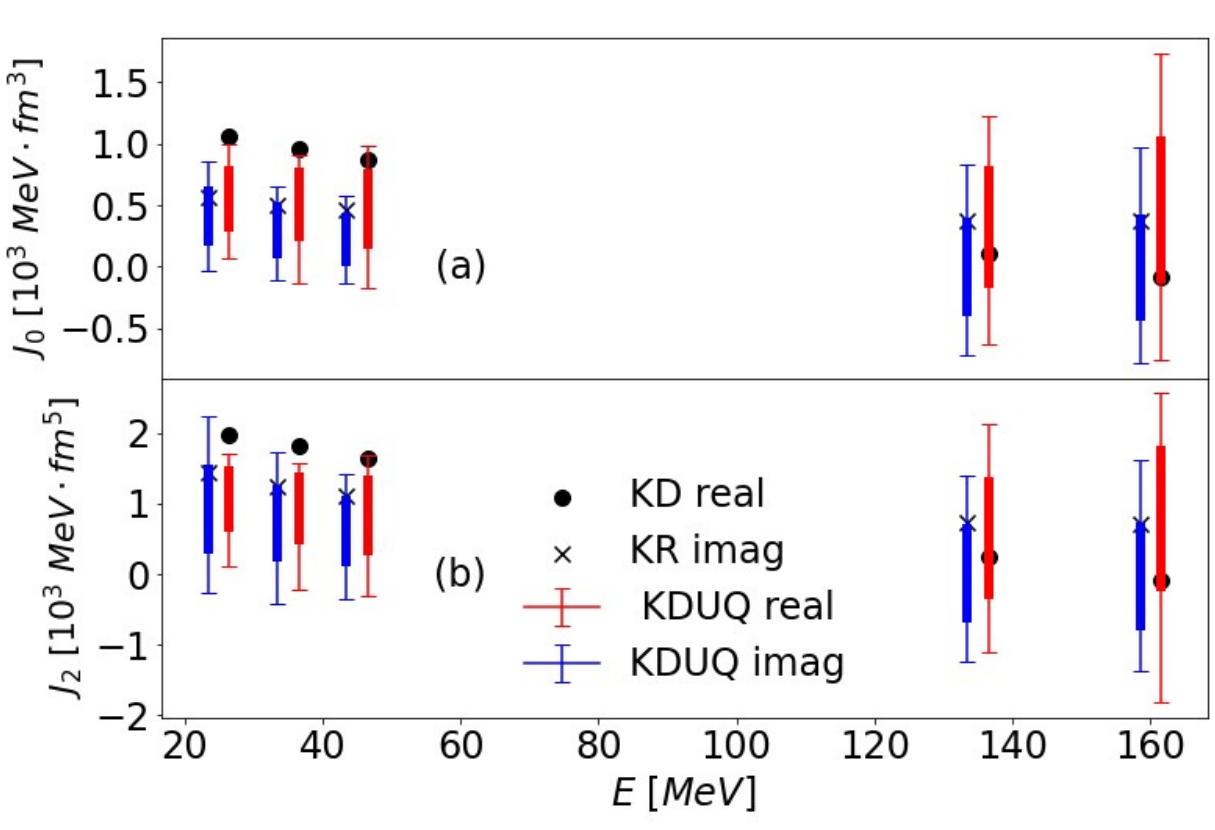}
        \caption{{Real and imaginary parts of the zeroth (top) and second (bottom) moments of the volume integrals of the transition operators  in the $s$-wave (Fig.~\ref{fig:fig3}) as a function of the beam energy. These moments are obtained with KD~\cite{KD}  (black points and crosses) and KDUQ~\cite{KDUQ} (red and blue error bars).   }}        \label{fig:fig4}
\end{figure}

{Finally, for the transition operators plotted in Fig.~\ref{fig:fig3}, we compute the first moments of the volume integrals, shown in Fig. \ref{fig:fig4}. We find that  the  uncertainties on the moments are large and the uncertainty of $J_2$ is of similar magnitude as that for $J_0$. These uncertainties increase with beam energy which 
imprints directly on the charge exchange cross sections. }

\subsection{Comparing two- and three-body models}

\begin{figure}
    \centering

        \includegraphics[trim=4cm 8.1cm 4cm 8.1cm,clip,width=\linewidth]{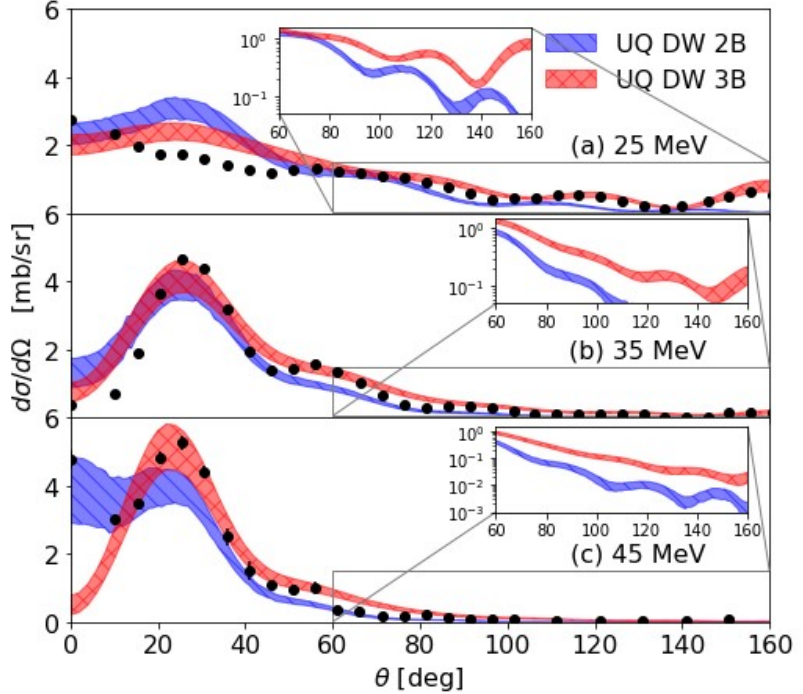}
  \includegraphics[trim=4cm 9.4cm 4cm 9.4cm,clip,width=\linewidth]{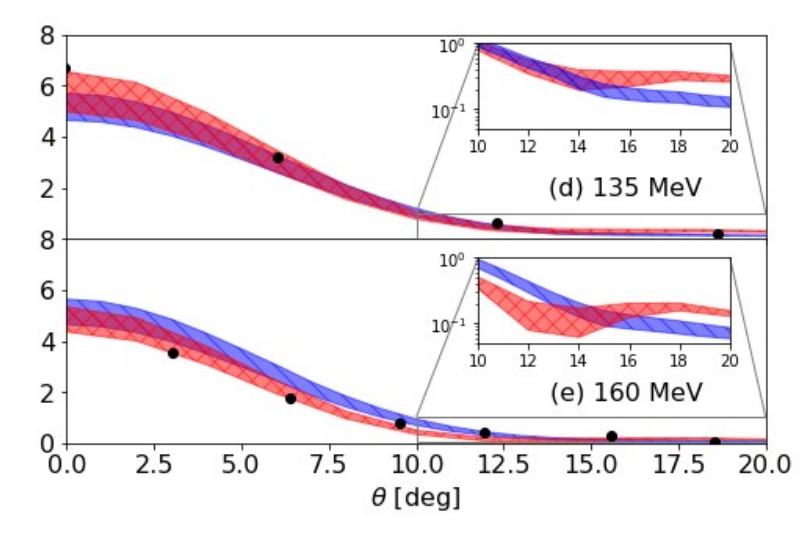}

        \caption{68\% credible intervals for charge-exchange cross section $^{48}$Ca$(p,n)^{48}$Sc($0^+$;IAS) at  (a) 25 MeV, (b) 35 MeV, (c)~45 MeV, (d) 135 MeV and (e) 160 MeV obtained within the two-body  (UQ-DW-2B,  blue forward hashmarks) and three-body (UQ-DW-3B, red crossed hashmarks) frameworks.
        These are obtained by propagating the KDUQ posteriors~\cite{KDUQ}  to just the scattering waves and  using the KD potentials~\cite{KD} for the operators. The inset shows the comparison of the theoretical predictions with the data at large angles on a logarithmic scale. }        \label{fig:fig2}
\end{figure}

In this section, we compare the simpler two-body approach with the more complex three-body model. Please note that in the latter case, uncertainties in $T_{NN}$ are not evaluated at present, thus, we include only the uncertainties emerging from the ambiguities in the optical potentials used to generate the initial and final distorted waves, which are needed in both the two- and three-body models. 
In Fig.~\ref{fig:fig2}, the results obtained when including the parametric uncertainties arising only from the optical potentials propagated to the distorted waves are shown, keeping the transition operator fixed. For the two-body calculations (UQ-DW-2B, blue band with  forward hashmarks), we consider the operator~\eqref{eq2} built from the original KD parametrization. For the three-body calculations (UQ-DW-3B, red   band with crossed hashmarks),  we  use the inputs as described in \ref{section:3B}. The bands displayed in Fig.~\ref{fig:fig2} correspond to the $1\sigma$ (68\%) credible intervals. The insets in Fig.~\ref{fig:fig2} correspond to the same angular distributions in log scale.

The  68\% credible intervals obtained in both frameworks agree overall reasonably well with the $^{48}$Ca$(p,n)^{48}$Sc($0^{+}$;IAS) data at all energies, indicating that both reaction models and operators are appropriate. There are small differences between the two calculations and the three-body calculations do perhaps slightly better in reproducing the experimental data, in particular at large angles. We note that some of the components of $T_{NN}$, including the central isospin ($\tau$) component, have a strong dependence on beam energy for $E_{p}\lesssim 100$~MeV~\cite{PhysRevC.24.1073,franey1985}. Since parametrizations of $T_{NN}$ were available at $E_{p}=$10, 20, 30, 40, 50, 100, 140, and 175 MeV, we used the ones closest to the experimental beam energies, and which reproduced the experimental data best. For $E_{p}= 25$ MeV and 35 MeV, the $T_{NN}$ from Refs.~\cite{CAPPUZZELLO2023103999,lenske2023} at 30 MeV was used. For the case of $E_{p}=45$ MeV, the available interactions from Refs.~\cite{CAPPUZZELLO2023103999,lenske2023} at 40 MeV and 50 MeV did about equally well, and the results for $T_{NN}$ at 40 MeV are shown in Fig.~\ref{fig:fig2}. When using the interaction at 50~MeV from Refs. \cite{PhysRevC.24.1073,franey1985}, the results were significantly worse.  For $E_{pp}=135$ and 160~MeV, the $T_{NN}$ parameterizations at 140~MeV and 175~MeV from Refs. \cite{PhysRevC.24.1073,franey1985} were used, respectively.       

The angular distributions vary strongly as a function of beam energy. Only at the highest beam energies are the distributions clearly forward peaked as expected for a low-momentum-transfer transition involving no change of angular momentum. At these higher energies, the cross sections are strongly dominated by the central isospin ($\tau$) component of $T_{NN}$ and the contributions from other (non-central) components of $T_{NN}$ are negligible. Therefore, at these energies, the operator used in the three-body model most closely resembles the operator used in the two-body model, and one expects the closest correspondence, which is consistent with the results shown in Fig.~\ref{fig:fig2}.

For beam energies below 50 MeV, it was observed that the non-central tensor component and, to a lesser degree, the spin-orbit component of $T_{NN}$ are important for obtaining good consistency with the data. 
At the lowest beam energy, the angular distribution at forward scattering angles is not well reproduced by the three-body theory. This could be an indication that the reaction process is more complex than assumed, and that in-medium effects and/or multi-step contributions are starting to play significant roles at this energy.

For both the two-body and three-body calculations the uncertainty bands obtained  in Fig.~\ref{fig:fig2}  are strongly suppressed compared to the one in Fig.~\ref{fig:fig1}. This can also be observed in Table~\ref{tab:table1} by comparing the relative uncertainties at the peak  obtained in the two-body model in the case of ‘‘Full UQ" and  ‘‘UQ-DW-2B". At all energies, the relative uncertainties stemming from the distorted waves are around 10-15\%.  This indicates that the two-body transition operator is responsible for the larger part of the uncertainties. 


\section{Conclusions}

In this work, we quantify uncertainties associated with charge-exchange reactions to isobaric analogue states deriving from the optical potentials. We use the recently developed KDUQ global parameterization \cite{KDUQ}, which is obtained from a Bayesian analysis of the original KD potential \cite{KD}. Pulling from the KDUQ parameter posterior distributions, we obtain credible intervals for the charge-exchange cross sections. As an illustrative example, we consider  reactions $^{48}$Ca$(p,n)^{48}$Sc($0^+$;IAS) for a wide range of beam energies $E=25-160$ MeV. We make predictions for the charge-exchange angular distributions using either the macroscopic two-body model or the more microscopic three-body approach, both based on 1-step DWBA. 

In the two-body model, we are able to fully quantify the interaction uncertainties by including the  uncertainties in the initial and final distorted waves as well as in the transition operator. The resulting uncertainties in the cross sections at the peak of the distributions span $60-100$\% . The credible intervals obtained for the charge-exchange cross sections are wider at the highest beam energies because the parameter posteriors for the imaginary part of KDUQ at the higher energies is broader. Note that, in KDUQ, the correlations between neutron and proton optical potentials are taken into account and, therefore, our estimated uncertainties are not artificially inflated as was the case in Ref.~\cite{whiteheadchex}. Our study shows that it is the uncertainty coming from the transition operator that dominates the uncertainty in the predicted charge-exchange cross sections. The uncertainties on the cross section from the distorted waves alone are less than about 15\%.   {
This work suggests that information
about the isovector part of the potential and the
symmetry energy that may be extracted from these cross sections will carry large parametric uncertainties. } 

We also compare directly the two-body and three-body model predictions for the charge-exchange angular distributions. To our knowledge, this is the first time such a comparison is done using the standard codes in experimental analyses. For the three-body model, we only include the uncertainties coming from the distorted waves, since the transition operator is obtained in a separate procedure for which no uncertainty quantification has been performed. If we only consider the uncertainties in the distorted waves, the angular distributions obtained within the two-body model are similar to those in the three body model, although the latter has a slight edge in describing the data.
For the example here chosen, the agreement with the data may suggest that, even for such a wide range of energies, 1-step DWBA is sufficient to describe the process, except perhaps for the data taken at the lowest beam energy of 25 MeV. 

An obvious way to reduce the uncertainties in predictions of charge-exchange is to  include charge-exchange data in the calibration of optical potential parameters. Such an approach is currently being implemented.

\begin{acknowledgements}
We thank Terri Poxon-Pearon for  sharing the CHEX code, Horst Lenske for sharing the $T_{NN}$ matrices used in the three-body model, Pablo Giuliani and Kyle Godbey for insightful discussions regarding the implementation of integration of KDUQ into CHEX,  Pawel Danielewicz and Kyle Beyer for feedback on the manuscript. This work was supported by the National Science Foundation, NSF-2209429 Windows on the Universe: Nuclear Astrophysics at FRIB. The work of F.M.N. was in part supported by the U.S. Department of Energy grant DE-SC0021422 and National Science Foundation CSSI program under award No. OAC-2004601 (BAND Collaboration). We gratefully acknowledge computational support from iCER at Michigan State University.

\end{acknowledgements}

\bibliographystyle{apsrev}
\bibliography{ChexSmith}

\end{document}